\documentstyle[]{mn}
\input psfig
\title[TMGS vs SKY]{Structure in the First Quadrant of the Galaxy: an Analysis of ``TMGS" Star Counts using the ``SKY" Model}
\author[P.~L. Hammersley  et al.]{P.~L. Hammersley$^{1}$, M. Cohen$^{3}$, F. Garz\'on$^{1,2}$, T. Mahoney$^{1}$, M. L\'opez-Corredoira$^{1}$ \\
$^1$ Instituto de Astrof\'\i sica de Canarias, E-38200  La Laguna,  Tenerife, Spain\\
$^2$ Departamento de Astrof\'\i sica, Universidad de La Laguna,  La Laguna,  Tenerife, Spain\\
$^3$Radio Astronomy Laboratory, 601 Campbell Hall, University of California, 
Berkeley, CA 94720}

\date{received date : accepted date}
\begin{document}

\maketitle

\begin{abstract}

We analyse the stellar content of almost 300 deg$^2$ of the sky close
to the Galactic plane by directly comparing the predictions of the
SKY model (Cohen 1994,1995; Wainscoat et al. 1992) with star counts
taken from the Two Micron Galactic plane Survey (TMGS: Garz\'on  et
al. 1993).  Through these comparisons we can examine discrepancies
between counts and model and thereby elicit an
understanding of Galactic structure.  Over the vast majority of areas in
which we have compared the TMGS data with SKY's predictions we find
very good accord; so good that we are able to remove the disc source
counts to highlight structure in the plane.  The exponential disc is
usually dominant, but by relying on SKY's predicted disc counts we have
been able to probe the molecular ring, spiral arms, and parts of the
bulge. The latter is clearly triaxial. We recognize a number of
off-plane dust clouds not readily included in models.  However, we find that,
whilst the simple exponential extinction function works well in the
outer Galaxy, closer than about 4 kpc to the Centre the extinction
drops dramatically. We also examine  the shape of the
luminosity function of the bulge and argue that the cores of all
spiral arms we have observed contain a significant population of
supergiants that provides an excess of bright source counts over those of a
simple model of the arms.  Analysis of one relatively isolated cut
through an arm near longitude 65 degrees categorically precludes any
possibility of a sech$^2 z$ stellar density function for the disc.
\end{abstract}


\section{Introduction}

Although  star counts have long been used for the determination of Galactic 
structure (see Paul 1993), 
there are still a  large number of parameters which are not  well determined.  
Reviews of star count  surveys can be found in
Bahcall (1986), Majewski (1993) and  Reid (1993).
Many visible studies have been directed towards the Galactic poles  
where extinction in the Galactic plane can largely  be ignored.
 Star counts, velocity distributions and 
metallicity  have been used
to determine and separate the various Galactic populations, such as the old disc, 
 young disc and halo. Detailed reviews are given, for example, by
 Gilmore, Wyse \& Kuijken (1989) and Freeman (1987), and 
references therein.
Hence, the vertical structure of the local Galaxy is well measured (although
there remain contentious issues such as the thick disc).
The horizontal structures are less well studied.  Robin, Cr\'ez\'e \& Mohan  
(1992) obtained deep visible star counts in the 
anticentre
region and concluded that the disc has a scale length of 2.5~kpc and a radial cutoff 
at about 14~kpc. 
 However, apart from  individual clear windows, visible star counts in the plane
 are limited to a maximum 
heliocentric distance of about 3~kpc (Schmidt-Kaler 1977), because of the
 high extinction.  A method of 
overcoming this,
at least in part, is  to observe in the infrared, where the extinction is
significantly lower. Infrared source counts  detect sources to a greater distance and
are far less affected by local dust, so the
measured distribution of sources is  closer to the true distribution.

\begin{table*}
\caption{Principal attributes of SKY.  For more details see WCVWS, Cohen (1994,1995),
and Cohen et al. (1994).}
\begin{tabular}{cccccc}
\hline
\hline
Geometric& No. of types& Radial scale& Source scale& Mag. range& Wavelength\\
components& of source& length& height& modelled& range\\
\hline
disc; bulge; arms, spurs, Gould's Belt; ring; halo & 87& 3.5 kpc& 90--325 pc& ($-$10,+30)& 0.14--35 $\mu$m\\
\hline
\hline
\end{tabular}
\end{table*}

\begin{table*}
\caption{Summary of regions of the sky analysed. Listed for each strip
are the declination and $l$ as the strip crosses the GP. Normally a
strip is identified by one of these two parameters. Also listed
are  the starting and ending points for each strip, the number of
useful areas of comparison with SKY, the area per bin, and the 
relevant figures 
in the Appendix for this strip. The area is important as it relates
observed TMGS counts in an area to counts per deg$^2$ (and the 
associated Poisson errors are calculable as shown in Figure 1).}
\begin{tabular}{ccccccl}
\hline
\hline
Dec. & Plane crossing & Start& End& No. useful& Typical& Fig. No. \\  
(J2000)&  {\it l}   & ({\it l},{\it b})& ({\it l},{\it b})& zones& area (deg$^2$)&     \\
\hline
    $-$30 & $-$1 & (6.54,$-$14.5)   & ($-$5.53,7.5)   & 45 & 1.9   & A1,B1,C1,D1 \\ 
    $-$22 & 7  & (15.007,$-$15.0) & (-2.223,14.5) & 61 & 2.0   & A2,B2,C2,D2 \\ 
    $-$11 & 21 & (27.495,$-$15.0) & (11.619,14.5) & 61 & 0.69  & A3,B3,C3,D3 \\ 
    $-$5  & 27 & (29.294,$-$4.5)  & (24.124,5.5)  & 20 & 0.18  & A4,B4,C4,D4 \\
    $-$1  & 31 & (38.994,$-$15.0) & (23.264,15.0) & 61 & 2.7   & A5,B5,C5,D5 \\ 
    +3  & 37 & (39.74,$-$5.5)   & (33.968,6.0)  & 24 & 0.25  & A6,B6,D6   \\ 
    +19 & 53 &(56.215,$-$5.0)   & (50.582,5.5)  & 22 & 0.18  & A6,B7,D7    \\ 
    +27 & 65 &(73.688,$-$5.5)   & (66.817,5.5)  & 61 & 2.0   & A7,B8,C6,D8 \\ 
    +32 & 70 &(75.837,$-$15.0)  & (56.995,15.5) & 60 & 0.56  & B9,D9    \\ 
\hline
\hline
\end{tabular}
\end{table*}

\begin{table*}
\caption{Summary of lines-of-sight covered. Listed for each strip is
the closest approach to the GC (assuming R$_\odot$=8500 pc) and the
features that are sampled. The question marks with the ring/bar
depend on the interpretation of the counts (see Sections 8,9). The Arm
crossings listed are those closest to the sun (Sct=Scutum,
Sgr=Sagittarius, and Per=Perseus arm). The $^*$ indicates that the
line of sight is almost tangential to the spiral arm.}
\begin{tabular}{ccclllll}
\hline
\hline
Dec. & Plane crossing & Closest distance & Bulge & Ring/bar & Sct & Sgr & Per \\  
    &  {\it l}& to the GC (pc) &  &   &  &  &   \\
\hline
    $-$30 &$ -$1 & 150  & Y & ? & Y   &  Y   &   \\ 
    $-$22 &  7 & 1050 & Y & ? & Y   &  Y   &   \\ 
    $-$11 & 21 & 3050 &   & Y & Y   &  Y   &   \\ 
    $-$5  & 27 & 3850 &   & Y & Y   &  Y   &   \\
    $-$1  & 31 & 4400 &   & ? & Y$^*$ &Y   &   \\ 
    +3  & 37 & 5100 &   &   &     &  Y   &   \\ 
    +19 & 53 & 6800 &   &   &     &  Y$^*$ &  \\ 
    +27 & 65 & 7700 &   &   &     &      & Y \\ 
    +32 & 70 & 8000 &   &   &     &      & Y \\ 
\hline
\hline
\end{tabular}
\end{table*}

  To date, infrared surveys of the Galactic plane capable of detecting 
point sources
have been limited. The largest IR survey to date was that
performed by the {\sl Infrared Astronomical Satellite},   $IRAS$, 
which mapped nearly the entire  sky at 12, 25, 60 and 100~$\mu$m. 
Much
work on Galactic structure has been enabled by $IRAS$ particularly on the
distribution of very late type stars (see Beichman,
1987, for a review). However, at large distances, $IRAS$ could detect only extremely 
luminous sources  (e.g. OH/IR stars) and was severely 
confusion-limited over much of the Galactic plane.
The Two Micron Sky Survey (Neugebauer \& Leighton 1969:
hereafter TMSS) also had a low limiting
magnitude ($m_K\approx 3$) and so could not detect sources in the inner Galaxy.
Other surveys with greater sensitivity have tended to be restricted to 
the Galactic Centre (hereafter GC) region. Those that have, at least in part,
examined the disc (e.g. Eaton et al. 1984; Kawara et al. 1982) have covered
 only 
very small areas of sky, typically a few hundred  arcmin$^{2}$. These surveys
ran the risk of sampling an area with unusually high extinction, 
as well as suffering from low source counts, and were unable
to show the large-scale distribution of sources.

Garz\'on et al. (1993) have described the  Two Micron Galactic
Survey (hereafter TMGS). This is a $K$--band  survey with an
approximate limit of  $m_K\sim$10 which represents a valuable addition to
our knowledge of the near--infrared Galaxy, deepening the broad sweep
coverage of the plane by a factor of almost 1000 compared with the
TMSS.  It fulfils an important transitional role, providing a
homogeneous data base  over a number of regions in the Galactic plane
at a time when both the hemispheric DENIS (Epchtein 1998) and all-sky
2MASS (Skrutskie 1998) three--band surveys are in routine operation.
These new surveys will go about 4--5 magnitudes fainter than TMGS and
offer simultaneous observations in $IJK^\prime$ or $JHK^\prime$.
Nevertheless, a $K$-band survey with a magnitude limit of  $m_K\sim$10
such as the TMGS can usefully probe the structure of our Galaxy.

Hammersley et al. (1994) and Calbet et al. (1995) explored
the morphology  of star counts from  the TMGS by heuristic methods, while
Hammersley et al.  (1995) deduced the displacement of the Sun
from the plane from the slight north--south asymmetry of counts
crossing the plane. However, these papers are based principally in
looking at the very obvious features seen in the the TMGS. In order to
understand the $K$ star counts more fully, a detailed model is required
which allows the  total counts to be broken down into counts from the
various Galactic geometric components in each magnitude range.  The
model that we have chosen to work with  is described by
 Cohen (1994), who has enhanced the ``SKY" model for the point source
sky originally described by Wainscoat et al. (1992: hereafter
WCVWS), adding molecular clouds for mid--infrared realism, extending
the Galactic structure to model the far--ultraviolet sky, and
displacing the Sun 15 pc north of the plane (Cohen 1995).

The purpose of this paper is to describe our use of the SKY model to
provide a formal basis for interpreting the TMGS source counts.  Our
eventual objective is to refine the model, guided by any discrepancies
that we find with the TMGS data base. To do this  we contrast the TMGS 
data from almost 600 patches of sky
(some 296 deg$^2$)  in the vicinity of the plane with the predictions
of the SKY model, to investigate the stellar content and probable
geometry of the Galactic plane.  The advantage of this approach is that
we can emphasize otherwise slight deviations from expectation, and
thereby define those regions in which the TMGS has uncovered substantive
evidence of structure contrary to that in a formal model.  We present
analyses with SKY of TMGS star counts along entire survey strips, from
10$^\circ$ to 30$^\circ$ in length, cutting the plane at a variety of
longitudes. We conclude that SKY provides a
very good overall description of $K$ counts in the plane, which is
chiefly dominated by the exponential disc, but that the TMGS offers important
insights into the character of structure very close to the Galactic
plane and the nature of the bulge. We find four basic styles of statistically 
significant deviation of
SKY's predictions from the observed counts.  One appears to arise from
a significant constituent of the ring, or arm, populations not
currently in SKY, and strongly confined in latitude close to
the plane in the inner Galaxy (longitudes
20$^\circ$--31$^\circ$).  A second type of behaviour suggests that the
luminosity function (hereafter LF) of the inner bulge is not that
implicitly represented by SKY.  The third set of departures from an
ideal model is caused by spiral arms that do not lie in the plane (cf.
WCVWS; Cohen 1994).  The remaining type of pattern is due both to
off--plane dark clouds not included in SKY's smooth analytic expression
of the extinction, and lines of sight through the Galaxy which have
unusually high or low distributed (rather than localized) extinction.

\section{The TMGS and the SKY model}
\subsection{The TMGS}
The TMGS  (Garz\'on et al. 1993) is  a collaborative project
between the  Instituto de Astrof\'\i sica de Canarias, Tenerife, and
Imperial College, London, to map large parts of the Galactic bulge and
plane. Observations were  made between 1988 and 1995 on  the 1.5-m
Carlos S\'anchez Telescope
 of the Observatorio del Teide, Tenerife.  Some 350 deg$^2$ of sky have
been mapped down to about 10 mag in $K$, resulting in a data set of
about 700,000  sources. The main  areas are in strips about
30$^\circ$ in RA (centred on the Galactic plane) by about 1.5$^\circ$
in declination. Most  TMGS sources lie within a few degrees of the
Galactic plane, the majority having no optical counterparts to at least
$m_V$=16, implying that they are sources deep within the Galaxy.  The
star counts presented  in this paper were made using all of the data in
each area.  Although the data were taken in strips in RA, the TMGS
catalogue has been converted to Galactic coordinates and the counts
 made using 1$^\circ$ bins in latitude.

\subsection{SKY: the Model}
The original SKY model described in WCVWS  incorporates six fully
detailed Galactic components: an exponential disc; a Bahcall--type
bulge; a de Vaucouleurs halo; a 4--armed spiral with 2 local spurs;
Gould's Belt and a circular molecular ring.  Each geometrical element
may contain up to 87 source categories, derived from detailed analyses
of the content of the $IRAS$ sky (Walker et al. 1989).  These
comprise 33 ``normal" stellar types; 42 types of AGB star, both oxygen-
and carbon-rich; 6 types of object that are distinct from others only
in their MIR high luminosity; and 6 types of exotica including H~II
regions, planetaries, and reflection nebulae.  Every category has its
own set of absolute magnitudes and dispersion in ten  hardwired
passbands, and an individual scale height and volume density in the
local solar neighbourhood.  Some sources are absent from some components
(the arms and ring are made rich in high--mass stars; the halo is 
deficient in these).

We have used version 4 of SKY as detailed by Cohen (1994,1995) and
Cohen et al. (1994), the principal attributes are summarized in Table 1. This  is different from WCVWS in that it has
the solar offset from the plane and reduced halo:disc ratio,  however,
we have not included the additional molecular cloud populations
essential to obtain a high-fidelity representation of the mid--infrared
sky.  This omission is justified because the clouds were inserted by
Cohen (1994) for $IRAS$ wavelengths of 12 and 25 $\mu $m, which are
essentially insensitive to extinction.  Therefore only the extra
sources were included and not the extra extinction through the clouds,
yet the extra populations associated with these clouds are surely
accompanied by extra extinction not accounted for within SKY.
 Cohen et al. (1994) demonstrated in the far--ultraviolet that
one could assign a value to the additional extinction encountered
through local molecular clouds using SKY.  Within the TMGS archive one
can clearly recognize obscuring clouds beyond which no 2-$\mu $m
sources are detected, although at $K$ only the dense cores cause a
significant effect on the counts and so the effect of the extinction is
generally limited to single TMGS zones. Consequently, without a
detailed prior knowledge of the additional extinction along the
specific TMGS lines of sight, it is preferable to model without
including these molecular complexes at wavelengths susceptible to
appreciable attenuation through clouds of $A_V$$>$10 mag. In this way
one can more readily identify those lines of sight that intercept
dark clouds at $K$. Furthermore, the $IRAS$ sources detected in the
clouds  had warm dust which makes the sources very luminous at 12
$\mu$m. This warm dust does not have as large an effect on the absolute
magnitudes at 2.2 $\mu$m, so the sources will be detected at $K$ but
will not be that much brighter than the sources without warm dust.
Hence they will be more difficult to see against the background of disc
or bulge sources.  This is the sole modification made to the published
description of SKY4.

\subsection{Why Confront SKY and the TMGS?}
The SKY model was originally built to reproduce the $IRAS$  12- and
25-$\mu$m source  counts. However, it has been successfully compared
with observations at many other wavelengths including the
far-ultraviolet (Cohen 1994;  Cohen et al. 1994).  The model has
already been compared with DENIS $K$ star counts in small regions
(Ruphy et al. 1997) for the magnitude range $m_K$=10--13.  However,
the comparison with the TMGS is possibly the most challenging for the
model because its magnitude range, $m_K$=5--9, probes the brightest
sources in the LFs of the geometric components such as the arms and
bulge. At $m_K$, these LFs are largely unknown, particularly for the
arms, but the contrast with the disc will be maximal so any discrepancy
with the model will readily be seen. Further, the areas to be studied
are long cuts through the arms and disc, providing nearby reference
regions out of the arms.  This, too, allows differences to be explored
in a manner that is impossible given only a single region. The final
important advantage is the relatively large area of sky covered which,
in many regions, affords meaningful star counts to be produced as
bright as $m_K$=5.

The fact that SKY produces good matches to $IRAS$ (Cohen 1995) and
visible stars counts towards the NGP (WCVWS) does not guarantee that
it will successfully match the TMGS star counts. The  $K$ LF
 is dominated  by photospheres whereas, for IRAS, the principal
populations detected at significant distances are all dust-enshrouded.
 Clearly, the $IRAS$ sources will be detected by the TMGS but are many
magnitudes fainter, greatly diminishing their importance. Thus, whilst
SKY may get the LF of the bulge correct for $IRAS$, this does not
necessarily imply correctness for the TMGS.  Garz\'on et al.
(1993) already noted that the Galaxy at 2.2 $\mu$m looks very different
from that in the visible because of extinction in the plane.
Furthermore, the principal sources detected by
 the TMGS are K4--M0 giants or M-supergiants, which are far less
 important in visible star counts when coupled with extinction.

\section{Confrontation of the Model by the Observations}
\noindent
In the Appendix (Figures A1--A7, D1--D9), we present TMGS counts along nine strips 
in the first quadrant.  These
 illustrate our interpretations of the counts by direct comparisons with 
the predictions of SKY.  Differential counts were scaled per square degree per
 magnitude, using entirely independent TMGS data in 0.2-mag wide bins. 

Table 2 summarizes for each strip: the J2000 declination; 
the longitude of the Galactic plane crossing; the bounding coordinates; 
the number of useful (i.e. with tolerable implicit Poisson errors) zones
we present in the relevant figure; the typical area sampled at each position along the strip
(this varies slightly along a strip); and the figure which illustrates 
the star counts and SKY predictions.  Table 3 amplifies this information
by indicating the closest approach of each line of sight to the Galactic
Centre and which (non-disc) geometric components contribute to the predicted
counts.

Our analyses depend on three basic analytical tools: differential star counts
(Figures A1--A7, D1--D9);  profiles of cumulative counts in a specific $K$-magnitude
bin as a function of latitude along each strip (Figures B1--B9); and similar 
profiles created after subtracting the disc component predicted by SKY from 
the observed counts (Figures C1--C9).  In both Figures B and C, we
represent the observations by solid lines and the model counts by dashed lines.

\begin{figure*}
 \centering
\vspace{24.5cm}
\includegraphics{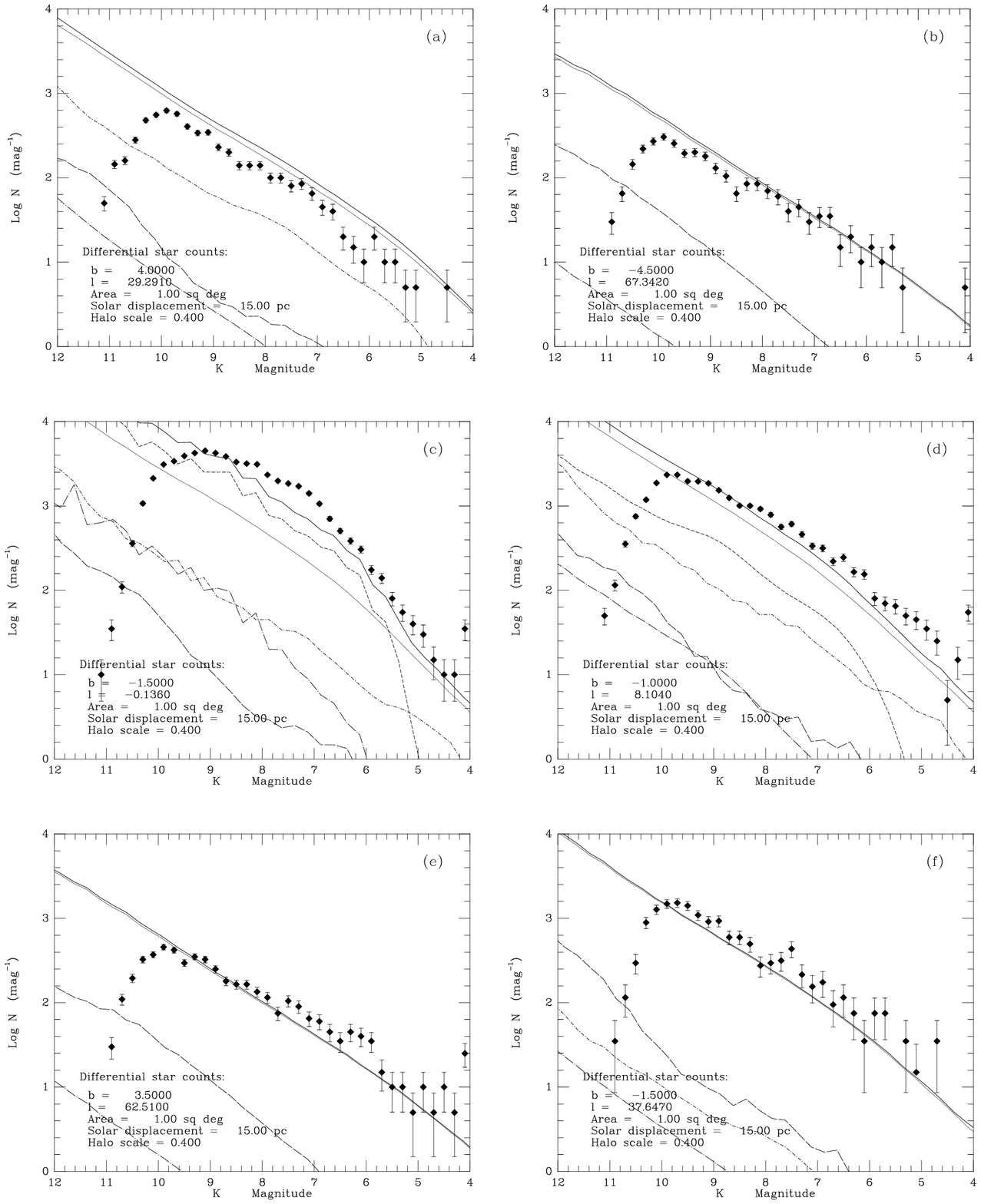}
\vspace{-2cm}

{{\bf Figure 1:}a-f }

 \label{f2a}

\end{figure*}
\begin{figure*}
 \centering
\vspace{17.5cm}
\hspace{-19cm}\includegraphics{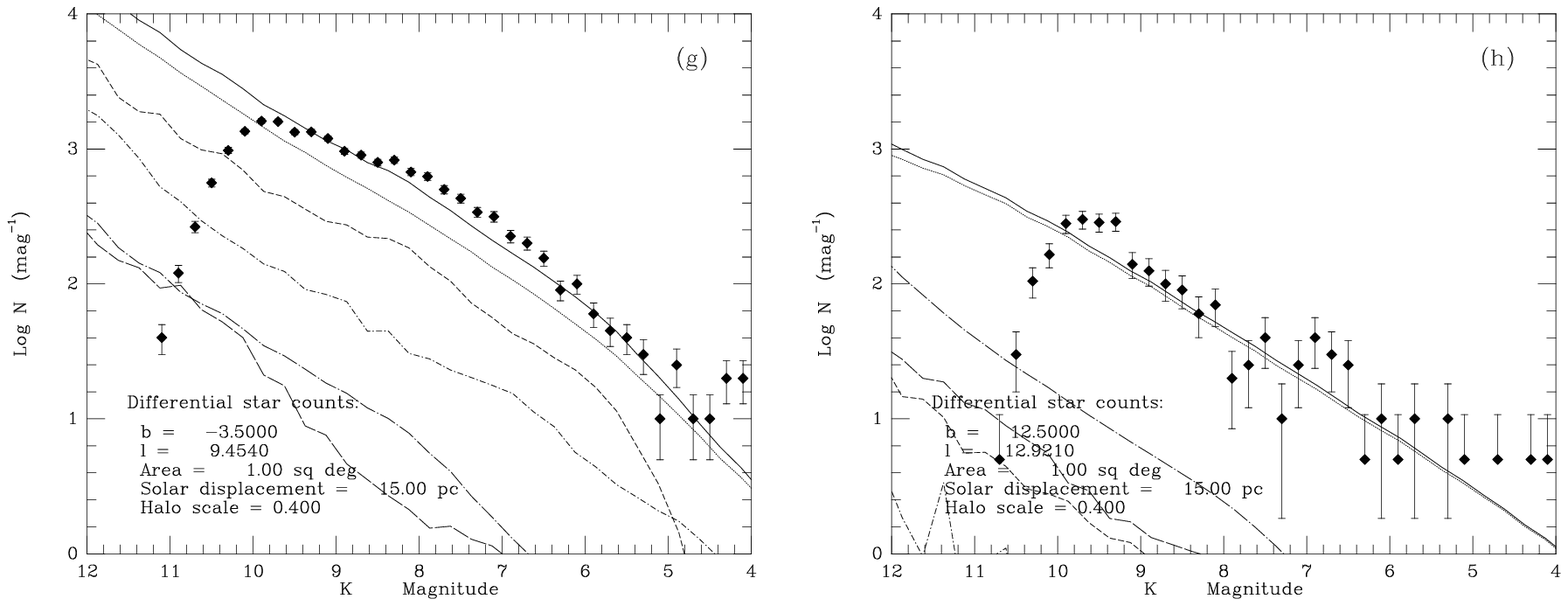}
\vspace{-10.1cm}

  \caption{g-h. Differential star counts  versus magnitude showing the 
detailed  comparisons between the TMGS and SKY for sample regions. The black 
diamonds are the TMGS points with Poisson error bars marked. The SKY geometric
 components are marked as follows: solid line -- total count; dotted or 
less heavy line -- disc; long dashes -- spiral arms, local spurs and Gould's
Belt; short dashes -- bulge; short dashes and dots -- molecular ring;
long dashes and dots -- halo }
 \label{f2b}

\end{figure*}

Figures 1(a)-(h) present eight representative fields, taken from the montages shown in
the Appendix, for which we explicitly plot the Poisson error bars appropriate
 to the
original counts (which were scaled by 5, to yield counts per mag from the 
0.2-mag binned data, and again, according to the actual area sampled, to yield
 counts per deg$^2$).
Figure 1 distinguishes the counts arising from separate
geometric components as follows: solid line -- total count; dotted or 
less heavy line -- disc; long dashes -- spiral arms, local spurs and Gould's
Belt; short dashes -- bulge; short dashes and dots -- molecular ring;
long dashes and dots -- halo.  The figures were chosen to be representative 
of the various regions covered as well as showing specific features in their
form which are discussed later in the paper.

As with all surveys, Malmquist bias can arise in TMGS and we
see its effects in Figure 1(h) where the counts in the final few faint
bins rise abruptly before the limit of completeness is attained.  Note
also that this is a region in which the actual source density is relatively
low.  In general, the direct comparison of SKY and TMGS counts indicates
that TMGS is complete to a magnitude between 9.2 and 9.8, depending on
the local surface density of sources.

\section{Extinction}
\subsection{Off-Plane Dark Clouds}
As with all previous versions of the model, SKY4 represents interstellar
extinction simply by the product of two exponentials; one radial, with
scale length of 3.5 kpc (matching that of the stars); one vertical, with
scale height of 100 pc (matching that of the youngest stars still 
associated with dark clouds).
This version of SKY does not include off-plane dark clouds so, if a TMGS 
strip runs through the core of such a cloud, the counts will be significantly
 decreased
when compared with the model (however, the greatly reduced extinction at $K$ 
compared with the visible implies it requires many magnitudes of visible 
extinction to cause a significant effect on the TMGS counts). 
 Figures A5,B5,D5 illustrates this phenomenon for 
the area centred on $l=29.3^\circ$, $b=4^\circ$ from the $l=31^\circ$ strip.
One sees that the TMGS counts fall well below the predictions
of SKY yet, at $b=2.5^\circ$ or $b=5^\circ$, the model and the TMGS are in good agreement. 
Figure 1(a) enlarges the plot for $b=4^\circ$, the deepest discrepancy, and adds the 
Poisson error bars.  This location corresponds to a known off-plane cloud which
clearly shows up in emission in longer-wavelength $COBE$  maps,  whereas the area 
is almost devoid of visible stars. In fact, the core of this cloud does not fill 
the TMGS bin at this location and sampling at higher resolution would 
decrease the TMGS counts even more at the centre of the cloud.
The only potential H~I counterpart (Weaver 1998, priv. com.) is a fairly uniformly
 dark patch (i.e. an absence of H~I) well-centred in an H~I emission ring, at an LSR 
velocity of $-$15 km s$^{-1}$, and about 1--1.5$^\circ$ in diameter. 
 Figure 1(b) shows another example,
this time of a more localized off-plane cloud, whose influence is recognizable
by an abrupt change in the observed source counts fainter than a specific
magnitude.  Here one sees an offset in source counts
by 0.2 dex for $m_K>$8.5.  Analysis of this line of sight by SKY indicates that
the dominant contributors at this magnitude are disc K0-3 giants, at a
probable distance of about 700 pc.  Consequently, we suggest that one encounters 
a cloud with incremental $A_V\sim$ 5 mag in this direction and at this distance.  
Again, H~I maps  (Weaver 1998, priv. comm.) show a sizeable void, 
bigger than that seen above (about 2$^\circ$ in diameter), well-centred on this TMGS
field, in an emission ring at an LSR velocity of +38 km/s.  In both cases, the 
presence of H~I emission with central voids suggests that the hydrogen in these clouds 
is largely molecular rather than atomic.

\subsection{On-Plane Dark Clouds and General Extinction in the Plane}
The TMGS reveals a number of ``on-plane'' ($|b|<2^\circ$) dark
clouds, particularly towards the inner Galaxy. Hammersley et al. (1996)
discussed the dark clouds which cause the major loss of stars at $l=-1^\circ$, 
$b=0^\circ$, and showed that most of the extinction comes
from individual dark clouds  close to the GC. However, there 
are individual clouds on, or very near, the plane in many TMGS areas
which locally cause a noticeable loss of counts ($l=7^\circ$, $b=1^\circ$ or 
$l=31^\circ$, $b=1^\circ$: Figures A2, B2, D2, A5, B5 and B5)

SKY includes only  the general extinction along the
line of sight. This is modeled as a double exponential which
continues all the way into the GC but  is normalized
in the solar neighbourhood. Figures B1-- B9 show the cumulative 
TMGS and SKY star counts (the latter separated into total and disc contributions)  
 for various strips, as a function of Galactic latitude, and the ratios 
of these totals at different limiting magnitudes. The effect of extinction 
on SKY's counts is clearly visible in these cuts as a dip in counts
at $b=0^\circ$. For lines of sight with $l>30^\circ$, the dip on the plane is 
small, a few percent of the total counts, and there is very good agreement with 
the TMGS star counts. In the inner Galaxy, however,  SKY predicts a
major loss of counts on the plane, principally due to extinction in front of 
the bulge sources (the difference between SKY's total and disc curves for 
$m_K>7$ and the $l=7^\circ$ and $-1^\circ$ strips is almost totally due to 
the bulge). The model does reasonably well
near $l=-1^\circ$, $b=0^\circ$ although, as already stated, most of the
extinction appears to come from identifiable dark clouds close to the GC,
 not represented in SKY. However,  SKY substantially over-predicts the
 extinction on  the plane at
$l=7^\circ$, $b=0^\circ$. The ratios of model-to-observed 
star counts for this region show a narrow peak which the original plot
indicates is due to extra extinction in  SKY. 

Hammersley et al. (1998) show that the extinction at the radius
of the molecular ring ($0.4R_\odot<R<0.5R_\odot$) is about a factor of three
above that predicted by a simple exponential model yet SKY's analytical
function for the extinction in the outer Galaxy (e.g.
R$>$ 0.5R$_\odot$) gives a reasonable representation for the distributed
extinction in the plane. Therefore, the only explanation for why SKY so
significantly overestimates the extinction in the line of sight
towards $l=7^\circ$, $b=0^\circ$, is that the extinction for $R>0.4R_\odot$ drops
to almost zero until a few hundred pc from the GC, where
the features associated with the Centre begin (e.g. the molecular disc). 
 The idea of there being a hole in the extinction  in the inner Galaxy is 
not new: the CO maps  clearly
show a significant fall-off in density inside the molecular ring
(Clemens, Sanders \& Scoville 1988; Combes 1991). Freudenreich (1998) also 
needed a hole in the
extinction in the inner Galaxy to match the DIRBE surface brightness
maps.

\section{The Exponential Disc for $b>5^\circ$}
SKY incorporates a single exponential disc with 87 categories of point source,
each with its own scale height, colours, and local space density.  The
scale length of all sources is set to 3.5 kpc.  SKY does not include a
thick disc for several reasons.  Firstly, it provides excellent fits to counts
at a wide variety of wavelengths without using a thick disc.  Secondly, the
proportion of disc counts contributed by the thick disc, even at optical
wavelengths where metallicities and kinematics may help separate these two
populations, is minimal ($<2\%$; e.g. Gilmore, Wyse \& Kuijken 1989). 
Thirdly, the stars of the thick disc
contribute negligibly in the infrared, far below the levels of Poisson errors
of the TMGS (and many other) data sets.

SKY separates young and old components
by assigning the young population (OBA dwarfs and supergiants) a 
significantly smaller scale height, 90 pc, than the
325 pc of the old populations.  Hence, the vast majority of the sources 
that SKY predicts that the TMGS  should see in the exponential disc are the old K- and M-giants
for regions well away from the bulge and  more than $5^\circ$ from the plane.  
The relevant figures are those for the
strips at $l=21^\circ,31^\circ$ and 65$^\circ$ (Figures A3, A5, A7, D3, D5 and D8). 
Figures B3, B5 and B8 show the associated cumulative counts for the TMGS
and the model for these regions, to $m_K=9$.

Both the form of the curves and the differential star 
counts show that SKY gives a good fit to the TMGS at all magnitudes in these 
longitudes, for  $|b|>5^\circ$.

In most areas of the sky, the disc dominates the star counts, making the 
remaining geometric components more difficult to identify and analyse. 
However, the accuracy of the fit of SKY'S disc to the TMGS counts makes it 
entirely reasonable to subtract the model's disc counts from the TMGS counts.
This  highlights the distribution of sources in the arms, ring, and bulge, 
if accomplished with care.

Alternatively, some argue that an exponential disc is no 
longer justifiable at the lowest latitudes in the Galaxy, and that the 
sech$^2z$ form might be 
physically more defensible. There are a number of papers comparing the sech$^2z$ 
and exponential in the discs of external galaxies (e.g. van de Kruit 1988; De Grijs,
 Peletier \& van de Kruit 1997). 
However, the results, even in the infrared, are complicated by extinction when close to the 
plane of the galaxy. The sech$^2z$ function 
asymptotically 
approaches an exponential as stellar densities fall far from the plane but,
for latitudes  within about 10$^\circ$ of the plane, it  fails utterly to
reproduce the observed counts.  We illustrate this failure in Figure 2.  The
$l$=65$^\circ$ strip is one in which we see only disc and arms, the
latter falling almost two orders of magnitude below the counts from the former.
Thus, it is an ideal location to experiment with a sech$^2z$ distribution
function. Figure 2 shows the cut through the plane at $m_K$=9 for two model 
discs for $l$=65$^\circ$,  $|b|\leq 15^\circ$, one is exponential in  height 
(this is a simplified model based on SKY but it is not the SKY model)  and the
second is identical except that it has a sech$^2z$ height distribution.  We have
 normalized the sech$^2z$ curve to the wings of
the cumulative latitude profile in the regions with $|b|>10^\circ$ as both 
functions work well, away from the plane.  Yet, for
all inner latitudes, the sech$^2z$ function delivers no more than half the
requisite counts.  This is equivalent to saying that, everywhere in the disc,
even in those locations where spiral arms and other components do not make
appreciable contributions to counts, more than 50\% of the observed stars
cannot come from the disc.  Were we to have carried out this test at other
TMGS positions in the plane, the excess of counts above the sech$^2z$ law
would be even larger than that shown in Figure 2.

We further note that the SKY model produces good fits to the DENIS $K$
star counts in the plane to $m_K$=13.5 (Ruphy et al. 1997),
significantly deeper than the TMGS. In regions away from the Galactic
bulge, the contribution from the disc will be at least an order of
magnitude greater than from any other component, e.g. the arms. If the
disc were not exponential, or at least very close to exponential,
all the way to the plane, then SKY would significantly 
overestimate the counts on the plane.

\begin{figure}
 \centering
 \vspace{11cm}
\includegraphics{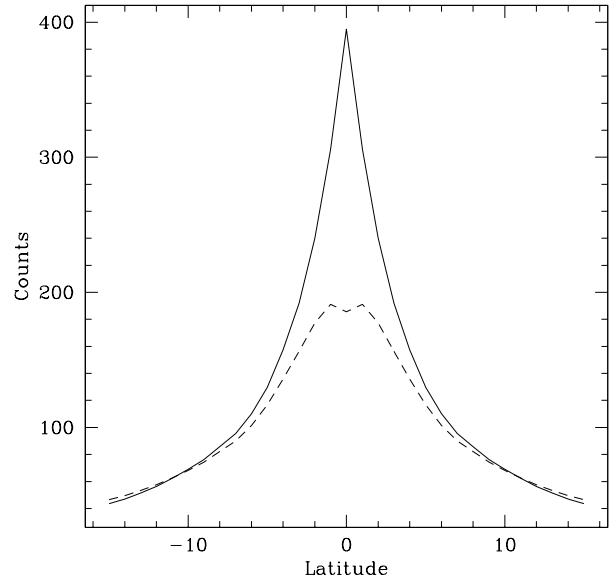}

 \vspace{-3cm}

 \caption{ Two simplified (not SKY) model cuts at $l$=65$^\circ$ for $m_K$=9. 
The solid line shows an exponential disc (as in SKY) the dashed is based on 
a  sech$^2z$. In both cases the SKY disc LF was used.
The two curves have been normalized at $b$=10$^\circ$.}

 \label{f2c}

\end{figure}

The $sech^2z$ function is strictly valid only for an isolated,
self-gravitating, isothermal system.  Self-consistent 
dynamical models of the disc (e.g. Haywood et al. 1997) show the 
dependence of the resulting z-distribution on star formation rate (SFR).  
For all but declining SFRs, the vertical distributions can mimic 
exponentials.  It is our suspicion that the $sech^2z$ function from a locally isothermal
disc achieved its popularity in the context of overall light
distributions across the planes of edge-on spirals at optical wavelengths
(e.g. van der Kruit \& Searle 1982).
In such a situation, heavy and patchy extinction militates against any
meaningful comparison between a $sech^2z$ law and observations and, far
from the planes of such galaxies, one cannot distinguish between $sech^2z$ and
exponential functions.  Perhaps the only redeeming feature of a $sech^2z$ law 
is that it offers a null second spatial derivative at zero latitude, yet 
this is precisely
the latitude range in which optical observations are invalid for the
purposes of making such comparisons.  Further, even in the
field of edge-on external galaxies, there is evidence that the infrared
light distributions follow exponential rather than $sech^2z$ functions
(e.g. Wainscoat {\it et al} 1989).

\section{The plane of the disc}

 For  $|b|<2^\circ$, at all longitudes, the TMGS has more counts on the plane than 
are  predicted by SKY. The profiles of counts (Figure B) show three 
distinct distributions at these latitudes:
\begin{itemize}
\item A distribution extending from about $|b|<4^\circ$ that is seen at all 
longitudes between $l=70^\circ$ and $-1^\circ$.
\item A spiked distribution about 1$^\circ$ wide centred on the plane at 
$l=21^\circ$ and $27^\circ$.
\item The bulge at $l=-1^\circ$ and $l=7^\circ$ (which we treat in section 9).
\end{itemize}

\subsection{The Spiral Arms}
SKY predicts that the 2-$\mu$m star counts should form a continuous
distribution from $b=15^\circ$ to $b=1^\circ$ with the only serious deviations
due to extinction in the plane.
However, the scans $l=21^\circ,31^\circ$ and 65$^\circ$ all show a 
distinct  transition at about $|b|=5^\circ$. (These can be seen both in the cuts
 [Figures B3, B5 and B8] and in the counts [Figures A3, A5, A7, D3, D5, D8 ].) 
The star counts for
$|b|>5^\circ$ approximate an exponential very well but, at $b=5^\circ$, it is clear that a 
second contribution is required with a far smaller scale height than typifies
the stars of the disc (i.e. an element rising more rapidly than the disc as
latitude decreases).  Figure 1(e)
represents this phenomenon from the $l$=65$^\circ$ strip, some 3.5$^\circ$
off the plane. 

A feature so close to the plane has to be due to a young component.
There are two possibilities to explain this distribution; a
young disc or spiral arms.  SKY already has a young component within 
its representation of the disc and these young stars (O, B and A) have a 
much smaller scale height than the older stars. Thus, if the extra sources
in the plane were young disc stars, this would imply an enhancement in
these populations.  Indeed, examination of Figures
B and C shows that, particularly at the brighter magnitudes, the
numbers of young sources would need to be increased by orders of
magnitude to reproduce the counts near the plane.  This would create 
a major distortion of the LF, clearly recognizable in the local vicinity.
Figure C shows the TMGS counts after subtracting the SKY disc.  Note 
that, for both $l=31^\circ$ and $l=65^\circ$, the width of the non-disc 
source distribution is independent of magnitude, suggesting that these
sources are grouped at a specific distance rather than spread along the 
line-of-sight.  For the above reasons, we prefer to attribute the excess 
of stars on the plane to the arms, rather than to a boost in the young disc.

The prediction from SKY is that
the  arms are always substantially below the disc counts,
although there are certainly locations in which the arms' contribution
is second only to that of the disc. However,
particularly at the brighter $m_K$ magnitudes ($m_K<7.5$) the TMGS finds
that the counts from the arms are at least as important as those from
the disc within 2$^\circ$ of the plane. The two major TMGS regions
(i.e. with sufficient  area to
 give reliable statistics at the brightest magnitudes)  that run only
 through spiral arms and through none of the other inner Galaxy
components are at
 $l=31^\circ$ and $l=65^\circ$. At $l=31^\circ$, $b=0^\circ$ at $m_K$=7,
 the TMGS has nearly three times SKY's  predicted  counts. At
$l=65^\circ$ at $m_K$=7,
 the excess on the plane is only 50\% but at $m_K$=5 the TMGS is again a
factor of 3 too high. This can be clearly seen in Figure B.  

 By $m_K$=9 in both the $l=31^\circ$ and 65$^\circ$ strips  the
 arm-to-disc ratio has
dropped significantly. In the range $8<K<9$ at $b=0^\circ$ there are some 331  
predicted disc sources compared with only 40 arm sources. At 
$l=31^\circ$ the loss of contrast is not as great (540 disc sources to 350 arm 
sources) probably because this line of sight is almost tangential 
to the Scutum arm, looking into the inner Galaxy, whereas the
 $l=65^\circ$ line of sight roughly maintains a
Galactocentric distance of R$_\odot$ for over 8 kpc.  

Another interesting conclusion can be reached by detailed comparison of
the star counts at $l=31^\circ$ and 65$^\circ$.  In Figures C5 and C6,
the dominant contribution to the disc-subtracted counts is from the
arms.  At $m_K=5$, we see roughly the same absolute number (within a
factor of 1.5) of luminous non-disc sources in both directions yet, by
the fainter magnitudes (e.g. $m_K=9$), the $l=31^\circ$ line of sight
shows about 6 times as many spiral arm sources as towards
$l=65^\circ$.  Thus, the LF of the spiral arms must be a function of
Galactocentric distance, rather than constant as in SKY4.  If this is
truly the case, one might suppose that metallicity was implicated.
There is a precedent for this argument in the observed confinement of
the latest WC-type Wolf-Rayet stars to the immediate environs of the
GC, a phenomenon attributed by Smith \& Maeder (1991) to the high
metallicity near the Centre which biases evolution of massive stars to
the formation of WC9 stars.

At $l=65^\circ$ (Figure C6), only the Perseus arm is crossed and at a
Galactocentric distance of about 8.5 kpc.  Here the excess is around 80
sources deg$^{-2}$ to $m_K$=8.  It is noticeable that the peak is not at
$b=0^\circ$; rather the centre of the distribution is shifted some
1.5$^\circ$ to positive latitude.  Also, the  observed counts  drawn
from the $l=37^\circ$ strip are parallel to, but above, SKY's
predictions at all magnitudes.  Both features are symptomatic of
directions in which the spiral arms do not lie in the formal Galactic
plane but arch above it. This same effect was detected in the IRAS
counts by WCVWS, who were able to match it to specific regions in the
H~I sky (e.g. Weaver 1974) where neutral gas lies almost 1 kpc off the
plane.  They cited, in particular, the $l=50^\circ$ to 90$^\circ$
region, which includes the off-plane peak of TMGS $K$ counts along the
$l=65^\circ$ strip.

\begin{figure}
 \centering
 \vspace{11cm}
\includegraphics{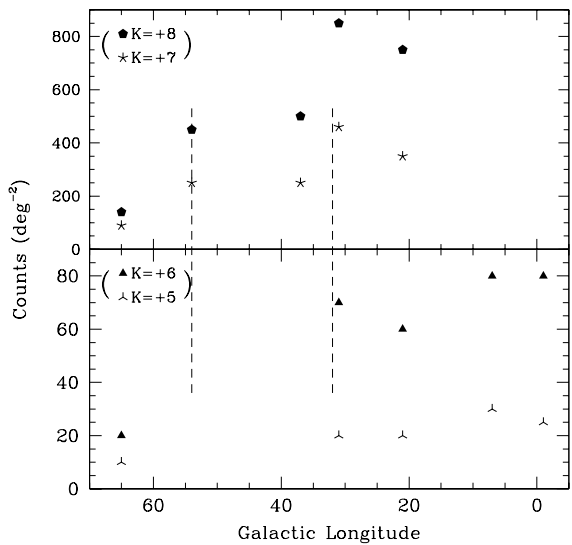}

 \vspace{-3cm}

 \caption{ The peak cumulative count at each TMGS plane crossing
after subtracting the
 SKY disc. The aim is to show those counts from the arms therefore the regions 
at 
giving a major contribution at fainter magnitudes.  Also at $l$=21$^\circ$  the 
counts are the estimated counts if  the ''spike'' (see below and Figure 3 
of  Hammersley et al. 1994) are ignored.}

 \label{f2d}

\end{figure}

Figure 3 shows the peak source count detected in each strip after the
SKY disc has been subtracted. Vertical dashed lines are the points
tangential to the
 Scutum and Sagittarius spiral arms. The aim is to show primarily the  arm sources on the plane, therefore, at $l=7^\circ$ and $l=-1^\circ$ only sources
to $m_K=6$ have been shown to avoid bulge sources. Also because of the spike (Section 6.2) the values at $l=21^\circ$  have been estimated as if the spike
were not there and the region at $l=27^\circ$ has been ignored. 
 The $l=31^\circ$ and 
 $l=37^\circ$  lines of sight are very close together, yet the
$l=31^\circ$  region has almost twice the number of arm sources as at
$l=37^\circ$.  The only significant difference is that the $l=31^\circ$
direction crosses the Sagittarius and Scutum arms whilst $l=37^\circ$
intercepts only the Sagittarius arm.  The extra sources to $m_K$=7 and 8
suggest that just under half the excess sources can be attributed to
the Scutum arm (i.e. about 350 sources deg$^{-2}$ to $m_K$=8 on the
plane).  The effect of the arms on the counts can be clearly seen in
all regions, in particular the influence of the number of arms which
cut the line of sight.  Directions crossing two arms have nearly twice
the counts of those crossing only one. Furthermore, the inner spiral
arms are far stronger than the arms at the distance of the solar circle
(cf.  the difference between $l$=65$^\circ$ and the other regions).
Another curious feature of Figure 3  is that at $m_K$=5 and 6 the number
of spiral arm sources does not alter significantly with longitude
between $l$=$-1^\circ$ and $l$=31$^\circ$.  Clearly there is a strong
interplay between the distance to the arm and the LF but one would
expect the  $l$=$-1^\circ$  and
 $l$=$7^\circ$  cuts to be significantly less than the  $l$=31$^\circ$
 which is almost tangential to the Scutum arm.

We attribute these excess sources observed near the plane  by the TMGS 
to the presence
of an additional population in the arms, not present in SKY4.
From the estimated distances to these spiral arms, this population
would have to be
of supergiant luminosity.  Consequently, the spiral arms clearly
provide an  important contribution to the $K$ star counts, particularly
towards the inner Galaxy.

\subsection{The ``Spikes''}
The most dramatic deviations from the SKY model anywhere in the area
covered here are the ``spikes'' on the plane at $l$=21$^\circ$ and
$l$=27$^\circ$.  Figures A3 and A4 portray the differential star counts
at $l$=21$^\circ$, $b$=0$^\circ$, and $l$=27$^\circ$, $b$=0$^\circ$,
where the excess is substantial at all magnitudes. If the SKY disc is
removed (Figures C3 and C4), then the TMGS has 6 times the counts of
the remaining SKY components at $l$=21$^\circ$. In the cuts (Figures B3
and B4), it can be seen that the distribution of the excess counts is
very narrow, about 1$^\circ$ wide, hence the term ``spike''. It is
significant that the spike only appears at $l$=21$^\circ$ and
$l$=27$^\circ$, there is no evidence for it at $l$=31$^\circ$ (or
larger longitudes). Neither is there evidence for the spike at
$l$=7$^\circ$ or $-$1$^\circ$ although the arms are clearly visible at
the brighter magnitudes in both strips so the absence of the spike
cannot be just a line of sight effect. This almost rules out the
feature being due to a continuous feature circling the inner Galaxy.
The implications of the spikes were discussed by Hammersley  et
al. (1994), who argued that these must be features associated with the
inner 3--4 kpc of the Galactic disc. The scale height must be around 40
pc, suggestive of  a very young population, and the $M_K$ must attain at least
$-11$ to $-13$. Mikami et al.  (1982)  suggested that there is 
a clustering of supergiants at $l$=27$^\circ$, $b$=0$^\circ$. Garz\'on et al. (1997) obtained visible spectra of some of the brightest
 TMGS sources with optical counterparts, proving that there are a very high 
number of supergiants in this region.  There are two possible explanations for what  causes
these spikes: bar or ring.

  In order to match the $IRAS$ 12- and 25-$\mu$m counts in this region
 WCVWS introduced another component, the molecular ring (Dame  et
al. 1987).  However, it is clear that, even with this representation
of the ring, SKY4 is well below  the TMGS counts.  The most significant
deviations are confined to  $\mid${\it b}$\mid$$\leq$1.5 $^\circ$ on
the {\it l}=21$^\circ$ strip, but persist to higher latitudes along the
{\it l}=27$^\circ$ strip ($\mid${\it b}$\mid$$\leq$3.5$^\circ$).  Note
that the slope of the observed counts in these regions more closely
parallels that of SKY's ring component than the steeply sloped disc
counts. Were we to posit that an additional stellar population attends
the ring, close to the plane, then this population must contribute some
5 times SKY's predicted ring counts near {\it l}=21$^\circ$  but almost
an order of magnitude more than SKY's ring near {\it l}=27$^\circ$.
The excess of sources above SKY's expectation is very similar on the
{\it l}=27$^\circ$ and 21$^\circ$ strips, but shows a conspicuous
additional bias toward very bright sources near {\it l}=27$^\circ$.

We note that Ruphy et al. (1997) describe an analysis of DENIS near-infrared
counts using a new version of SKY, with a non-uniform elliptical ring that
replaces the old circular molecular ring in SKY4.  In a later paper we plan to 
exercise SKY5 (after implementing all the changes discussed in the present
paper) to investigate what TMGS counts tell us about the geometry of this
ring.  In brief, DENIS counts probe the lower-density portions of this ring
and help to constrain the overall geometry of the ellipse.  But TMGS offers
direct lines of sight across the first quadrant dense ansa included
in SKY5.  Given sufficient TMGS strips, and new deeper $JHK$ images, the
density enhancements around this new ring should be definable. The
inclusion of these star-rich ansae provides SKY with a measure of asymmetry
(triaxiality) not present in earlier versions, and can be used to provide a
more formal examination of the bar.  However, we defer this discussion to
a future paper using new data and further developments of SKY.

Hammersley et al. (1994) suggested that such spikes could be
associated with a possible bar distribution on the grounds that, where
the bar interacts with the disc, there is a high probability of strong
star formation. Note this ``bar'' is not the triaxial bulge, often
called a bar, but rather is a  stick-like bar with a half-length of
nearly 4 kpc and an angle of 75$^\circ$ to the GC-sun line. Neither can
the star formation  be related to the triaxial bulge  somehow
triggering star formation in the inner disc (as suggested by
Freudenreich 1998) as the  geometrical constraints alone make this
nearly impossible.  Hammersley {\ it et al.} (1994) also examined the
$COBE$ 2.2-$\mu$m surface brightness maps and argued that such a bar far
more simply and naturally predicts the observed features (both images
and star counts), whilst the form of a  ring would have to be contrived
using {\it ad hoc} solutions to make it fit.

It is beyond the scope of the current paper to engage in a discussion
of ring  versus bar.  To make any significant advance, new data
are required. However, it should be noted that many external bars have
strong ansae of star formation at either ends and so the discussions in
the previous two paragraphs are not mutually exclusive.

\section{The Bulge}

 Two TMGS strips run through the bulge: those at  $l$=$-1^\circ$
and $l$=7$^\circ$. However, when examining these two regions, one must
remember that the strips cut the plane at an appreciable angle so there is a 
significant variation in $l$ as well as $b$, which is important this close to the GC.
These are the most complex regions to analyse because disc, arms,  
and bulge contribute to the counts.  The first step is to subtract the
SKY disc to highlight the bulge population.  This is entirely reasonable
because SKY gives a good fit to the exponential disc, and because of the high 
bulge-to-disc ratio fainter than $m_K$=7 (e.g. near the plane, there are
twice as many bulge stars as disc stars).

 Figures B1 and B2 show the cuts for the two strips; Figures C1 and C2 show the residuals after 
subtracting SKY's disc from the TMGS counts. At $m_K$=5 and 6 the residuals in 
the TMGS are clearly due to spiral arm sources because they are as narrow in 
spatial extent, and as large in size, as the residuals at $l$=31$^\circ$.
However, at $m_K$=6, SKY's predicted residuals clearly show 
a wider distribution than seen in the TMGS, due to the predicted bulge sources.
 Even at $l$=7$^\circ$  the SKY residual is far wider than seen in the TMGS,
yet by $m_K$=9 the residuals are far closer together.
The explanation for these discrepancies is that SKY's bulge LF 
 has too many of the most luminous sources. A quantitative 
analysis of the $K$ LF, based on the TMGS, is presented
in L\'opez-Corredoira et al. (1997,1999) and shows that  SKY's bulge LF
 starts 1 to 2 magnitudes too bright at $K$. Consequently, it is planned
 to implement a new LF for the bulge and to re-examine this issue in the future.

\begin{figure*}
 \centering
\vspace{22.5cm}
\includegraphics{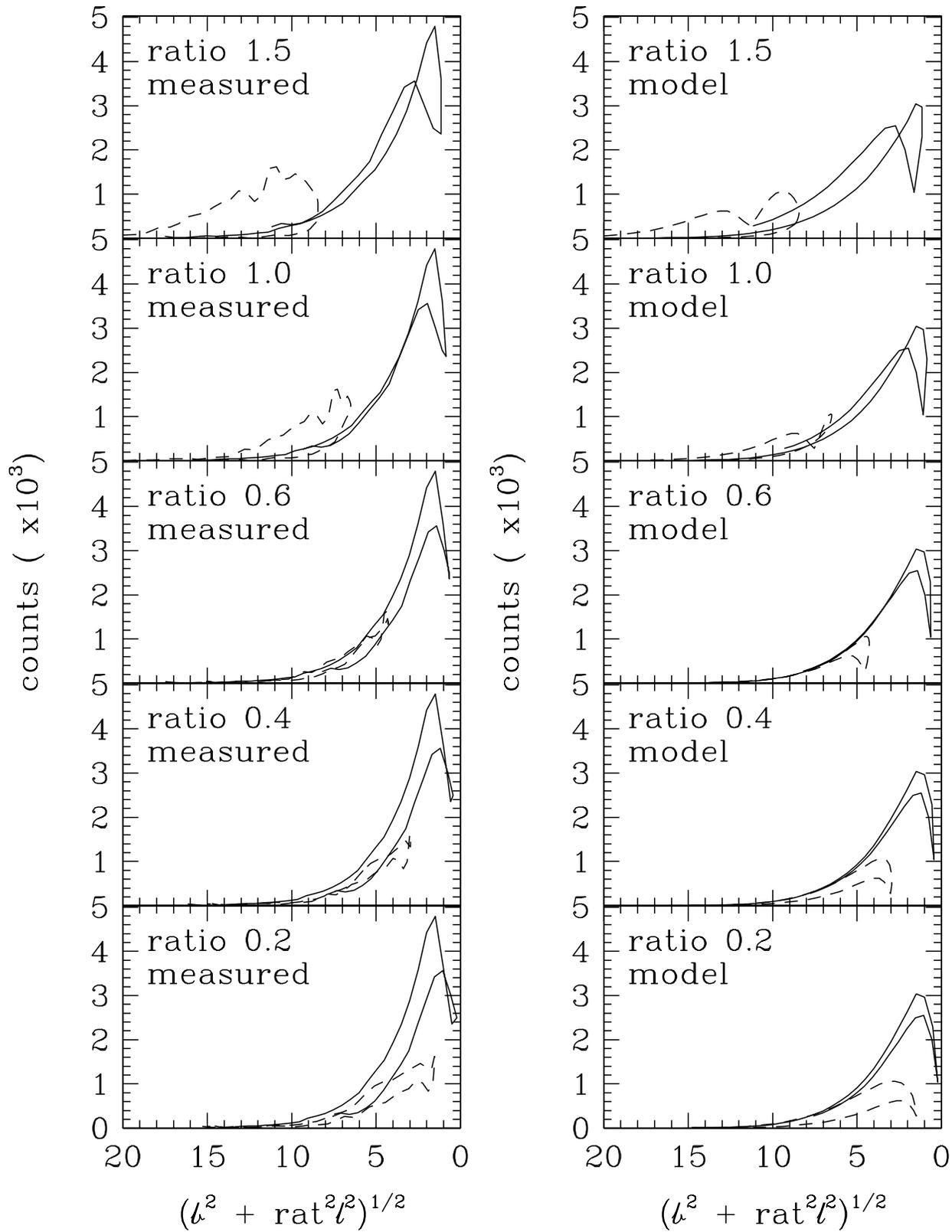}
\vspace{-0.1cm}

\caption{The left column shows the TMGS counts after subtracting the 
SKY disc plotted against  $  ( rat^2 l^2 + b^2 ) ^{0.5}  $  
  for various ratios ($rat$). 
The right column shows SKY total $-$ SKY disc plotted in the same manner. 
The strips are for 
$l$=$-1^\circ$ (solid line) and $l$=7$^\circ$ (dashed line).}

\label{f4b}

\end{figure*}

The SKY bulge is axially symmetric, with a  projected axial ratio of
0.6. Therefore, if one were to draw ellipses of constant predicted bulge
star counts, then these would have a ratio of major-to-minor
axis of 0.6;  i.e. all positions on the locus $\sqrt{0.6^2l^2+b^2}$ 
would have the same star counts. The bulge is the only component which has this type 
of projected spatial distribution and so plotting this function is an unambiguous test of
 whether the TMGS is actually detecting bulge sources. It should be noted that as the function  $\sqrt{0.6^2l^2+b^2}$ is always positive this means that the
line for a TMGS strip will double back on itself as it crosses the plane. 
  Figure 4 shows SKY's bulge counts plotted against    
$\sqrt{ rat^2l^2+b^2}$ (right column) together with the TMGS strips (left column)
 through $l=7^\circ$  (dashed line) and $l=-1^\circ$ (solid line) for various ratios,
 $rat$, between 1.5 and 0.2. 
In both cases this is  after subtracting the SKY disc and for $m_K$=9 so that
the bulge will dominate all the remaining components. 
More than a few degrees from the GC, the best fit for the bulge (i.e. the 
lines at the same position most nearly overlie each other is for 
a projected axial ratio of 0.6).  There is some evidence from 
the central few degrees (Figure 4, $l$=$-1^\circ$) that the fit is better 
for an axial ratio
of 1.0, but this is consistent with a triaxial bulge (as will be discussed below), which would  depress the portion from negative longitudes.

\begin{figure*}
 \centering
\vspace{22.5cm}
\includegraphics{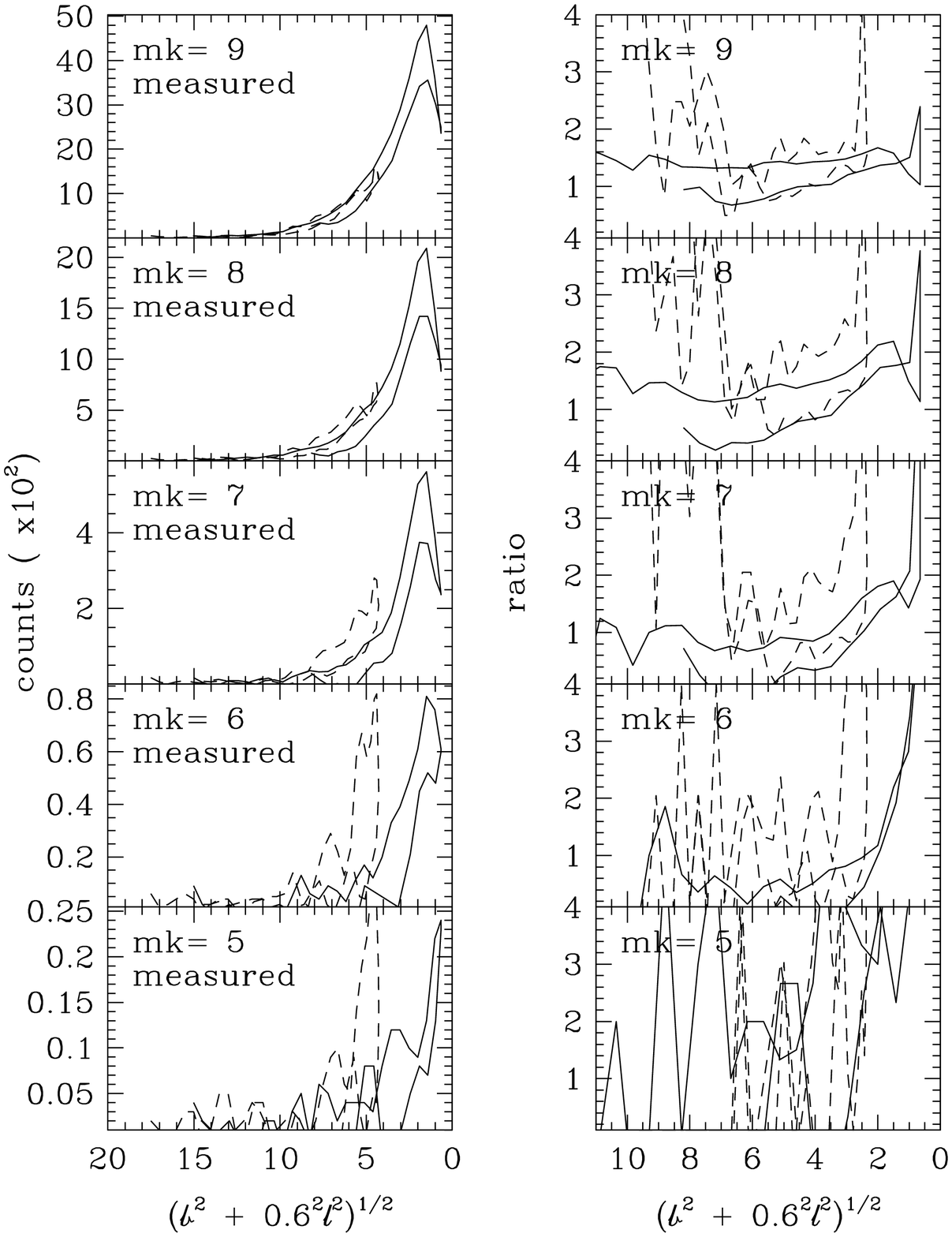}
\vspace{-0.1cm}

\caption{The left column shows the TMGS counts after subtracting the 
SKY disc plotted against  $  ( 0.6^2 l^2 + b^2 ) ^{0.5}  $  
  for various ratios $K$ magnitudes. 
The right column shows the ratio of (TMGS $-$ SKY disc) to  (SKY total $-$ SKY disc ) also plotted against  $  ( 0.6^2 l^2 + b^2 ) ^{0.5}  $.
The strips are for 
$l$=$-1^\circ$ (solid line) and $l$=7$^\circ$ (dashed line).}

\label{f4a}

\end{figure*}

  Figure 5  shows the same type of plot as Figure 4, this time varying
the $K$-magnitude at a fixed ratio of 0.6. The left  column shows the
TMGS whilst the right  column shows the ratio of (TMGS--SKY disc) to
(SKY total--SKY disc).  At $m_K$=5, where there is no bulge component,
there are two separate peaks caused by the spiral arm component in the
plane.  Spiral arms clearly will not follow the bulge locus. At $m_K$=6,
SKY (as discussed above), predicts that there should be a bulge-like
component but the TMGS plots show that the only significant population
is that of the arms. For this reason, apart from the plane crossings,
the TMGS is below  the counts  predicted by the model. $m_K$=7 is a
transitional magnitude where the bulge begins to become important,
although in the plane there is still a significant arm component. On
average, the TMGS residuals are below the predicted SKY residuals;
i.e. the TMGS bulge counts are below those of SKY.  On average at
$m_K$=8 and 9 away from the plane, the ratio of SKY to TMGS is about 1
suggesting that by these magnitudes the SKY LF is close to that
observed by the TMGS. Note that the peak in the ratio where  the TMGS strips
cut the plane (where the curves for each strip double back on
themselves)  at $m_K$=8  and 9 is  caused by the model overestimating the
extinction in the inner Galaxy (see Section 4.2).

  SKY predicts that the bulge distribution in the two cuts should be
asymmetric due to the change in $l$ as well as $b$. However, a
comparison of the TMGS residuals, taking account of the discrepancy in
the LF, shows that for $b<0^\circ$ ($l>0^\circ$ by comparison with the
point at $b=0^\circ$) the TMGS is invariably above SKY, whereas for
$b>0^\circ$ ($l<0^\circ$) the TMGS is well below SKY (Figures B1, B2, C1 and C2). It should be noted that
the effect is very obvious particularly at $m_K$=7 and 8 and so we are
not discussing some subtle feature in the counts.
 Initially it was
supposed that this was due to the extinction by major dark cloud
systems above the Galactic plane in this direction. However, recent
studies have shown that in general there is insufficient extinction in
these clouds to account for the observed effect we see (Mahoney 1998).
Typically, the additional $A_V$ above the plane, as opposed to below
it, is 1 mag (i.e. an extra $A_K$ of about 0.1 mag).  The asymmetry
seen in the TMGS would require a difference of extinction in $K$ of at
least 0.5 mag, which is impossible to attribute to extinction alone.
Figure 5 shows that at $l$=$-1^\circ$ (the solid line) the difference
between the TMGS/SKY ratio for  $b<0^\circ$ ($l>0^\circ$, the upper
part of the solid line) and $b>0^\circ$ ($l<0^\circ$) increases with
distance from the Galactic plane (between 2 and 7 on the x-axis).
However, the total extinction, as well as the difference in extinction above
and below the plane, must decrease away from $b=0^\circ$. Hence, if the
effect were due to extinction both the upper and lower parts of the
solid line would come together with increasing distance from
$b=0^\circ$, and not diverge as is seen.  Furthermore, if the
differential extinction above and below the Galactic plane were
significant, it  would clearly show in the $COBE$ 2.2-$\mu$m surface
brightness maps, which is not the case (see Freudenreich 1998).

   A closer analysis of the difference between SKY and the TMGS shows
that,  at $m_K$=8 and 9  (where the LFs are in closer agreement), the TMGS
is above SKY for $l>0^\circ$ and below SKY for $l<0^\circ$.
Furthermore, at $m_K$=7, where the asymmetry in the TMGS is the greatest,
the distribution is more bulge-like for $b<0^\circ$ (i.e. greater $l$)
than $b>0^\circ$, where the distribution is almost narrow enough to be
due solely to spiral arms, particularly in the $l=-1^\circ$ strip.  A
number of papers have argued that the bulge is triaxial with the near
end in  the first quadrant at an angle of 12$^\circ$ to 30$^\circ$ to
the GC-Sun line of sight (Binney et al. 1991: Dwek et al.
1995: Freudenreich 1997: L\'opez-Corredoira et al. 1997). This would on average  make the bulge sources  at  $l>0^\circ$ closer than at $l<0^\circ$.  If the
bulge were triaxial, we would expect asymmetries in the TMGS star
counts. The sources from the closer part of the bulge ($l>0^\circ$)
would appear at a
 significantly brighter magnitude than those with $l<0^\circ$.  A rough
estimate suggests a magnitude difference as high as 0.5 mag. The steep
rise in the bulge LF for the brightest sources means the counts
 cut on very sharply around $m_K$=6.5, but then rise less steeply  from
$m_K>$7.5. Thus  we would expect to see a very asymmetric distribution in
the $m_K$=7 cut but a far more symmetric one at $m_K$=9, precisely as seen
in the TMGS counts. Similarly, the ratios shown in Figure 5 have the
features that one would expect when ratioing star counts from a
triaxial bulge with those from an axisymmetric model; i.e. the more
positive the longitude, the more the counts rise above the predicted
values
 and the more negative the longitude, the more the counts fall below
 the predictions.  $K$ star counts give access to the LF so we  suggest
that an analysis of the bulge based on star counts is likely to be far
more sensitive than one based solely on surface brightness maps because
it avoids the ambiguities in the size and orientation of the triaxial
bulge inherent in an analysis of its surface brightness (Zhao 1997).
Recently L\'opez-Corredoira et al. (1997,1999) have  analysed the
TMGS star counts by inverting the fundamental star count equation -
very different from the qualitative analysis presented here - and have
found that the bulge is triaxial with an angle to the GC-Sun line of
12$^\circ$.

For the innermost strip ($l=-1^\circ$) almost sampling the GC, the latitude 
range in which there is statistically significant deviation from SKY that 
cannot arise from off--plane dark clouds, is confined to 
$\mid${\it b}$\mid$$\leq$2.0$^\circ$ (Figures A1, B1 and D1).  From these plots, it 
is apparent that, if the bulge component were simply enhanced by factor of 2,
the total predicted curve would much more satisfactorily match the counts.
This bears directly on the bulge LF implicit
 within SKY.  One can see from Figure 1(g) how it is possible, quite
directly, to probe this LF  because the observed counts are
well-matched by SKY except in the restricted range $8.4>m_K>6.2$. TMGS counts 
exceed SKY's here by about 50\%
at most, but this is naturally attributable to the shape of the bulge LF
which declines abruptly at bright $K$ magnitudes.  We note, however, that there is some
interplay between changes in the density function and changes in the bulge LF.

\section{ Conclusions}
\noindent
We conclude that, overall, SKY provides a very reliable simulator of the
TMGS Galaxy.  Only rarely are the predictions wrong by as much as 50\% 
and, in all
these cases, there is a fairly obvious change that can be implemented in
SKY to match the observations by more realistically representing the
astrophysical situation.  Consequently, the wide-area coverage
offered by TMGS provides substantive constraints on the model.  This
potential interaction between SKY and the real Galaxy is extremely valuable
in the development of a better model and in better understanding the 
structures in the inner Galaxy and their relative importance to $K$ star counts.

Specific changes that we plan to investigate in more detail and to implement
in a new version of SKY are: the inclusion of an additional population of
supergiants in the cores of the spiral arms; a modified LF for the bulge; a triaxial
bulge;
more non-coplanar arms and plane than represented by WCVWS; and
the non-uniformly dense elliptical ring.

Perhaps the single most important result is that the star counts  at 2.2
$\mu$m can  clearly be  modelled using simple analytical functions to describe
the 4 Galactic components of disc, arms, bulge and ring. 

\section{ Acknowledgments}
The  Carlos S\'anchez Telescope is operated on the island of Tenerife by 
the Instituto de Astrof\'\i sica 
de Canarias in the Spanish Observatorio del Teide of the Instituto de 
Astrof\'\i sica  de Canarias.

{}

\appendix

\section{The plots showing the comparison between TMGS and SKY}
The appendix presents the majority of the the plots showing
the detailed comparison between TMGS and SKY. From these plots it
is possible to see how the relative importance of the various Galactic
components  vary with magnitude and position.

 Figures A1 to A7 (9 plots)  illustrate our interpretations of the
counts by direct comparisons with  the predictions of SKY. The plots
are for a positions every few degrees along each scan (the full data is
presented in figure D). Differential counts were scaled per square
degree per magnitude, using entirely independent TMGS data in 0.2-mag
wide bins.  The counts arising from the separate geometric components
are  distinguished as follows: solid line -- total count; dotted or
less heavy line -- disc; long dashes -- spiral arms, local spurs and
Gould's Belt; short dashes -- bulge; short dashes and dots -- molecular
ring; long dashes and dots -- halo. The position for each area is given
in each plot.  See section 3 and Tables 2 and 3 for more details.

Figures B1 to B9 show in the left column the cumulative star counts
from the TMGS for each magnitude between $m_K$=5 and 9 plotted as the
solid line. The longitude at which the strip crosses the plane is given
in the upper box.
 The long dashed line shows the total  prediction from SKY and the
dotted line  the counts from the SKY disc alone. In the right column is
shown the ratio of TMGS counts to SKY total counts for each magnitude.

  Figures C1 to C6  show the cumulative counts after
subtracting the SKY disc from each magnitude between $m_K$=5 and 9.  The
longitude at which the strip crosses the plane is given in the upper
box. The solid line is TMGS -- SKY disc, The long dashed line is SKY
total -- SKY disc and the dotted line is the SKY bulge.

 Figures D1 to D9 (33 plots)  illustrate our interpretations of the
counts by direct comparisons with  the predictions of SKY every half a 
degree in latiude along the full TMGS scans. 
Differential counts were scaled per square degree per magnitude, using
entirely independent TMGS data in 0.2-mag wide bins.  The counts
arising from the separate geometric components are  distinguished as
follows: solid line -- total count; dotted or less heavy line -- disc;
long dashes -- spiral arms, local spurs and Gould's Belt; short dashes
-- bulge; short dashes and dots -- molecular ring; long dashes and dots
-- halo. The position for each area is given in each plot.  See section
3 and Tables 2 and 3 for more details. Due to the number of plots these 
are only available electronically.

\end{document}